# Efficient and Accurate Simulations of Vibrational and Electronic Spectra with Symmetry-Preserving Neural Network Models for Tensorial Properties

Yaolong Zhang[1,2], Sheng Ye[1,3], Jinxiao Zhang[1,3], Ce Hu[1,2], Jun Jiang[1,3], Bin Jiang[1,2*]

[1]*Hefei National Laboratory for Physical Science at the Microscale, Department of Chemical Physics, University of Science and Technology of China, Hefei, Anhui 230026, China*

[2]*Key Laboratory of Surface and Interface Chemistry and Energy Catalysis of Anhui Higher Education Institutes, University of Science and Technology of China, Hefei, Anhui 230026, China*

[3]*Chinese Academy of Sciences Center for Excellence in Nanoscience, University of Science and Technology of China, Hefei, Anhui 230026, China*

*: corresponding author: bjiangch@ustc.edu.cn

## Abstract

Machine learning has revolutionized the high-dimensional representations for molecular properties such as potential energy. However, there are scarce machine learning models targeting tensorial properties, which are rotationally covariant. Here, we propose tensorial neural network (NN) models to learn both tensorial response and transition properties, in which atomic coordinate vectors are multiplied with scalar NN outputs or their derivatives to preserve the rotationally covariant symmetry. This strategy keeps structural descriptors symmetry invariant so that the resulting tensorial NN models are as efficient as their scalar counterparts. We validate the performance and universality of this approach by learning response properties of water oligomers and liquid water, and transition dipole moment of a model structural unit of proteins. Machine learned tensorial models have enabled efficient simulations of vibrational spectra of liquid water and ultraviolet spectra of realistic proteins, promising feasible and accurate spectroscopic simulations for biomolecules and materials.

## TOC graphic

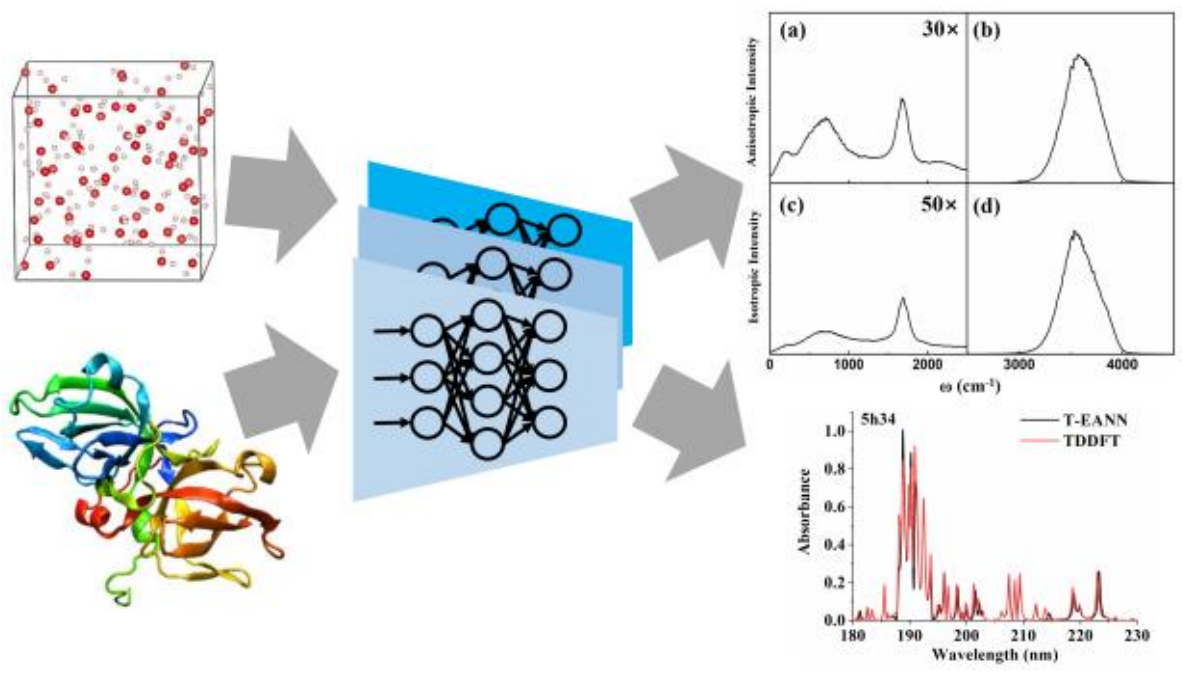

## I. Introduction

Machine learning (ML) techniques have shown great promise in solving challenging problems in physics, chemistry and material science[1]. Although various ML methods are well-established in computer science, one common problem is that they do not automatically recognize any intrinsic symmetry of any data. To describe a given physical quantity precisely, it is therefore necessary to symmetrize the ML representation in terms of translation, rotation, and permutation operation. By designing various symmetry-invariant descriptors as the input of nonlinear ML methods[2-11] or as the basis functions for linear regression[12-13], there have been quite successful ML applications in representing scalar property, *e.g.* the potential energy, which is invariant with respect to symmetry operations on the molecular configuration[14]. These ML-based potential energy surfaces (PESs) provide analytical and continuous atomic forces, thus making molecular dynamics simulations orders of magnitude faster than first-principles calculations.

However, much fewer studies have focused on the ML representation of tensorial molecular properties, *e.g.* permanent/transition dipole moment (PDM/TDM) and polarizability tensor. These tensorial properties are generally more difficult to learn, as they contain multiple coordinate-dependent components that are covariant when the system is rotated. One may bypass this challenge by manually aligning the molecules to a local reference frame[15-22]. This common practice is however less well-defined in heavily-distorted structures or dissociable systems and could cause discontinuity in the boundary of the reference frame. With kernel-based regression, Ceriotti and coworkers generalized covariant kernels to account for the rotational symmetry of the response tensorial properties[23-25]. Christensen *et al.* instead applied the response operators to kernel functions mimicking the tensorial response of potential energy to an electric field[26]. In comparison, the nonlinearity of conventional neural networks (NNs) would scramble the covariant symmetry embedded in structural descriptors, unless one directly makes the activation of each neuron covariant to rotations[27].

For PDM, it is well-known that one can simply learn the atomic charge multiplied with the corresponding coordinate vector, reducing the ML model to an efficient scalar analogue[28-33]. However, it is less well-noticed that this simple approach is not directly extendable to TDM, which describes the *transition* (or change) of charge distributions in two states, as shown below. It is neither applicable to the polarizability tensor. In this Letter, we extend this simple NN model to describe the TDM and polarizability tensor in an efficient way. The core idea is to construct the desired tensorial form by multiplying virtual NN outputs and/or their partial derivatives with atomic coordinate vectors, while keeping structural descriptors symmetry-invariant. We demonstrate that the proposed tensorial NN models yield accurate predictions for a large number of *ab initio* PDM, TDM and polarizability data in various benchmark systems. Furthermore, the NN-predicted TDM surfaces for a peptide fragment are used to faithfully reproduce the ab initio based electronic spectra of two realistic proteins with about six orders of magnitude speedup.

## II. Method

Let us go through the PDM ( $\vec{\boldsymbol{\mu}}$ ) case quickly, which corresponds classically to the separated charge ($q$) multiplied by the vector pointing from the negative to the positive charge ( $\vec{\mathbf{r}}$ ), *i.e.* $\vec{\boldsymbol{\mu}}=q\vec{\mathbf{r}}$ . For a given system with $N$ atoms, PDM can be readily obtained by the sum of atomic contributions[28],

$$\vec{\boldsymbol{\mu}}=\sum_{i=1}^{N} q_i \vec{\mathbf{r}}_i \,, \tag{1}$$

where $q_i$ is the atomic effective charge of $i$th atom (not physically meaningful) and $\vec{\mathbf{r}}_i = (x_i, y_i, z_i)^T$ is the corresponding atomic coordinate vector originated from the center of mass. Note that $q_i$ is a scalar quantity with the same symmetry of atomic energy that can be easily fit in an atomistic NN framework[29-30], which immediately warrants the translational and permutational invariance and the rotational covariance of PDM (see Fig. S1). Indeed, Bowman and coworkers[28] have pioneered the use of permutationally invariant polynomials (PIPs) to fit atomic charges and construct dipole moment surfaces for many molecular systems, but one needs a proper adaption of a global basis like PIPs which can be derived from the invariant theory[34] (see Refs. 12 and 13 for reviews of this PIP approach).

It may not be so obvious that this approach is inapplicable to TDM. The quantum description of PDM corresponds to the expectation of the dipole operator in terms of a given electronic wavefunction $\psi_i$ , namely, $\vec{\boldsymbol{\mu}}=\langle \psi_i | q\vec{\mathbf{r}} | \psi_i \rangle$ . In this regard, PDM relies only on the charge distribution of this electronic state. However, TDM associates with the transition between the two different states, $\vec{\boldsymbol{\mu}}_{\mathrm{T}}=\langle \psi_f | q\vec{\mathbf{r}} | \psi_i \rangle$ , which is well-known to be affected by the relative phase of the initial and final electronic states as well as the transition type (*i.e.* the change of charge distribution upon transition). The phase problem can be mitigated by a phase correction procedure comparing the overlaps between wavefunctions of neighboring configurations when generating training data[20]. However, the orientation of the TDM vector is dependent on the molecular orbitals involved in the transition, which is not at all taken into account in Eq. (1). For example, for a triatomic molecule lying in the *yz* plane with a perpendicular transition, the TDM vector is nonzero only in its $\mu_x$ component. However, the $x_i$ component in each atomic coordinate vector is zero, vanishing the right side of Eq. (1). As a consequence, the aforementioned atomic charge model would completely fail in such a simple case, giving rise to unavoidably large fitting errors. To solve the problem, we introduce two vectors in the same way as the PDM vector, namely,

$$\vec{\boldsymbol{\mu}}_{\mathrm{T}}^{j}=\sum_{i=1}^{N} q_i^j \vec{\mathbf{r}}_i \qquad (j=1,\,2), \tag{2}$$

where $q_i^j$ ($j$=1, 2) can be obtained by two different outputs of the same atomic NN and the directions of $\vec{\boldsymbol{\mu}}_{\mathrm{T}}^{1}$ and $\vec{\boldsymbol{\mu}}_{\mathrm{T}}^{2}$ are automatically determined by NNs (see Fig. S1). As long as $q_i^1$ and $q_i^2$ are not accidentally identical, $\vec{\boldsymbol{\mu}}_{\mathrm{T}}^{1}$ and $\vec{\boldsymbol{\mu}}_{\mathrm{T}}^{2}$ will define a specific plane (*e.g.*, the molecular plane for a planar geometry) and their cross product will give rise to a third vector perpendicular to this plane,

$$\vec{\boldsymbol{\mu}}_{\mathrm{T}}^{3} = \sum_{i=1}^{N} q_i^3 (\vec{\boldsymbol{\mu}}_{\mathrm{T}}^{1} \times \vec{\boldsymbol{\mu}}_{\mathrm{T}}^{2}), \tag{3}$$

where $q_i^3$ is given by another output of the same atomic NN that determines the magnitude of $\vec{\boldsymbol{\mu}}_{\mathrm{T}}^{3}$. These three fundamental vectors can be then linearly combined, namely $\vec{\boldsymbol{\mu}}_{\mathrm{T}}^{\mathrm{NN}} = \vec{\boldsymbol{\mu}}_{\mathrm{T}}^{1}+\vec{\boldsymbol{\mu}}_{\mathrm{T}}^{2}+\vec{\boldsymbol{\mu}}_{\mathrm{T}}^{3}$, to mimic a TDM vector that is not necessarily restricted in the molecular plane, with the correct rotational covariance (see Fig. S1). We keep using the symbol $q_i^j$ for consistency, which contains no physical meaning here. Since Eqs. (2-3) take advantage of the same atomistic NN structure with multiple outputs, the NN models for TDM and PDM are comparably efficient as their counterpart for PES. We note that $q_i^j$ ($j$=1-3) can be also fit, as the effective charge in Eq. (1), by many other robust ML method, such as PIPs and Gaussian process.

Next, we will follow this concept to design the tensorial NN model for molecular polarizability (**α**). Let us recall that **α** is the second-order response of potential energy ($V$) to an electric field $\vec{\mathbf{E}}$,

$$\boldsymbol{\alpha} = \frac{\partial \vec{\boldsymbol{\mu}}_{ind}}{\partial \vec{\mathbf{E}}} = \frac{\partial^2 V}{\partial \vec{\mathbf{E}} \partial (\vec{\mathbf{E}})^{\mathrm{T}}}. \tag{4}$$

where $\vec{\boldsymbol{\mu}}_{ind}$ is the induced dipole moment. Apparently, **α** is a second-rank tensor and a 3×3 symmetric matrix, which is translationally and permutationally invariant but covariant with respect to rotation. A convenient way to construct a 3×3 symmetric matrix is taking the product of a 3×$M$ matrix ($M \geq 3$) and its transpose. To this end, we can design an effective induced dipole-like NN structure similar to that in Eq. (2) with multiple ($M$) outputs,

$$\vec{\boldsymbol{\mu}}_{ind}^{j} = \sum_{i=1}^{N} q_i^j \vec{\mathbf{r}}_i \qquad (j = 1,\ 2,\ ...,\ M), \tag{5}$$

which has the same rotational covariance as the dipole moment. Alternatively, we can also calculate the partial derivatives of the virtual quantity generated by the NN model with respect to atomic coordinates, leading to a 3×$N$ matrix,

$$\mathbf{D}^j = \sum_{i=1}^{N} \frac{\partial q_i}{\partial \vec{\mathbf{r}}_j} \qquad (j = 1,\ 2,\ ...\ ,\ N), \tag{6}$$

Multiplying either $\boldsymbol{\mu}_{ind}$ or **D** matrix by its transpose gives us the required 3×3 matrix,

$$\boldsymbol{\alpha}^{\mathrm{NN1}} = \mathbf{D}(\mathbf{D})^{\mathrm{T}} \quad \text{or} \quad \boldsymbol{\alpha}^{\mathrm{NN1'}} = \boldsymbol{\mu}_{ind}(\boldsymbol{\mu}_{ind})^{\mathrm{T}} \tag{7}$$

We choose to work with the **D** matrix in this work, which is found to produce more accurate results (see SI for details). However, with either choice, $\boldsymbol{\alpha}^{\mathrm{NN1}}$ is a semidefinite matrix by construction, while molecular polarizability itself is not necessarily semidefinite. We thus create another symmetric matrix $\boldsymbol{\alpha}^{\mathrm{NN2}}$ in the following way,

$$\boldsymbol{\alpha}^{\mathrm{NN2}} = \mathbf{r}(\mathbf{D})^{\mathrm{T}} + \mathbf{D}\mathbf{r}^{\mathrm{T}}, \tag{8}$$

which is obviously not semidefinite. Furthermore, it is important to note that both $\boldsymbol{\alpha}^{\mathrm{NN1}}$ and $\boldsymbol{\alpha}^{\mathrm{NN2}}$ become a rank-deficient matrix when the molecular geometry is planar, while the molecular polarizability tensor is not. The simplest way to correct this is to incorporate a scalar matrix $\boldsymbol{\alpha}^{\mathrm{NN3}}$, whose diagonal element can be optimized by a very simple NN. Combining these three terms yields the full representation of the NN-based polarizability tensor,

$$\boldsymbol{\alpha}^{\mathrm{NN}} = \boldsymbol{\alpha}^{\mathrm{NN1}} + \boldsymbol{\alpha}^{\mathrm{NN2}} + \boldsymbol{\alpha}^{\mathrm{NN3}}, \tag{9}$$

that fulfills the symmetry of $\boldsymbol{\alpha}$ (see Fig. S1).

In practice, any atomistic NN methods proposed to represent scalar quantities can be readily adapted within the formalism discussed above, and importantly, one needs no modification of the symmetry-invariant descriptors. In this work, we generalize our recently proposed embedded atom neural network (EANN) model to representing these tensorial properties. The accuracy and efficiency of the scalar EANN model have been demonstrated in our recent publication.[9] The key advantage of the EANN approach is that it scales linearly with respect to the number of neighbor atoms and the total number of atoms in the system. More details on the implementation of the tensorial EANN (T-EANN) model are given in the Supporting Information (SI).

## III. Results and Discussion

We first apply the T-EANN model to water-related systems, for which there have been ab initio data available for comparison in the work of the symmetry-adapted Gaussian approximation potential (SA-GAP) method developed by Ceriotti and coworkers[23]. For each system, one thousand ab initio data of $\vec{\boldsymbol{\mu}}$ and $\boldsymbol{\alpha}$ were randomly divided into training and test sets with a 50:50 ratio. Due to the widespread numerical ranges of ab initio $\vec{\boldsymbol{\mu}}$ ($\boldsymbol{\alpha}$) values, we compare in Table I the root mean squared errors (RMSEs) relative to intrinsic standard deviation (RRMSEs) of the testing samples, with those obtained by SA-GAP[23]. Fig. S2 shows more explicitly scatter plots of ab initio data and NN predictions and the absolute RMSEs. The T-EANN models accurately represent both quantities in all systems, yielding RRMSEs of $\boldsymbol{\mu}$ ($\boldsymbol{\alpha}$) as 0.02% (0.02%), 6.6% (4.2%), 1.3% (0.3%), and 16% (2.2%) for $H_2O$, $(H_2O)_2$, and $H_5O_2^+$, and liquid water, respectively. The RRMSEs for water dimer and liquid water seem somewhat larger than others, due possibly to the more diverse data points (*e.g.,* the former includes both the bound state and dissociation continua of two monomers and the latter evolves in a huge configuration space, see also their numerical ranges in Fig. S2), neither conditions are beneficial to NN-based models. But overall, as shown in Fig. S2, T-EANN predictions correlate quite well with ab initio values in all systems. It is known that kernel-based models typically require fewer data than NN-based ones to reach the same level of accuracy[11, 35]. It is thus encouraging that, despite its simplicity, our T-EANN model generally outperforms the SA-GAP model with such a small amount of data.

The prowess of the T-EANN model is further supported by the simulation of Raman spectra based on the machine learned polarizability tensor surface, taking liquid water as an example. As there were no matching data of potential energy for liquid water reported in Ref. 23, we choose to fit another set of freely available data[36] using the EANN approach[37]. The resulting EANN potential was used in the classical molecular dynamics simulations, by which the Fourier transform of autocorrelation functions of polarizability tensors yields isotropic and anisotropic Raman spectra shown in Fig. 1 (see the SI for more details). Despite the inconsistency between energy and polarizability, we find that the simulated Raman spectra of liquid water successfully capture the strongest O-H stretch peaks at ~3520 $cm^{-1}$, much weaker peaks for H-O-H bending around 1680 $cm^{-1}$ and the broader librational band below 1000 $cm^{-1}$. Importantly, we estimate that the T-EANN polarizability model is $1.5\times10^4$ faster than direct DFT calculation. This result validates that accurate and efficient spectroscopic simulations would directly benefit from our machine learned polarizabilities.

Our T-EANN model for polarizability differs in spirit from the SA-GAP model which accounts for the covariant symmetry in terms of the tensorial smooth overlap of atomic positions (λ-SOAP) kernels and the regression is linear. Ceriotti and coworkers[25] also discussed the framework of combining the λ-SOAP representation with NNs. Because NNs consist of nonlinear mapping functions in hidden layers, the tensorial λ-SOAP features have to be incorporated to the linear output layer of a scalar SOAP-based NNs where the output serves as a multiplier for the tensorial features[25]. In other words, the hidden layers of NNs virtually output linear expansion coefficients for the tensorial λ-SOAP descriptors only. This is completely different from the T-EANN representation implemented here, in which the virtual outputs of EANN are differentiated with respect to nuclear coordinates, similar in spirit to the ML models for the high-dimensional electronic friction tensor of an adsorbate at a metal surface[38] and the nonadiabatic coupling vector[39] between two electronic states. The extension of the T-EANN model to higher-rank tensors is less obvious but not impossible, as one can in principle construct the target high-rank tensor by the direct product of low-rank ones. We also note that Sommers *et al.* reported a tensorial deep NN model for polarizability tensor during the preparation of our manuscript[40]. The rotational covariance is incorporated in a similar way as we construct $\boldsymbol{\alpha}^{\mathrm{NN1}}$, namely by multiplying a coordinate-dependent $\mathbf{T}$ matrix with its transpose, with an additional NN-based diagonal matrix inserted in between (Eq. 12 in Ref. [40]). This treatment renders the resultant tensor unnecessarily semidefinite, but in our opinion, could still suffer from the singularity issue in planar geometries as we have discussed above. This is however not crucial in their application in condensed phase systems.

We discuss next the T-EANN model for TDM. There were scarce ML models representing global TDM surfaces[19-21]. However, to our best knowledge, none of them rigorously considered the covariant symmetry of the TDM vector. To show the critical role of the symmetry, we train a T-EANN model with the TDM data of the N-methylacetamide (NMA) molecule (see Fig. 2a). The nπ* and ππ* excitations of NMA (see Fig. 2b) have been extensively used to model the ultraviolet (UV) spectra of the amide group of the protein backbone[19, 41]. Our training set consists of the TDM data for n$\pi$* and $\pi\pi$* transitions of NMA at the time-dependent DFT (TDDFT) and PBE0/cc-pVDZ level with the Gaussian 16 package[42], for ~50000 NMA configurations extracted from 1000 different protein backbones at room temperature from the RCSB Protein Data Bank (PDB)[43]. Phase corrections have been carefully done by evaluating the wavefunction overlaps of neighboring configurations along the trajectories[20].

In Figs. 2c-f shows the correlation diagrams between the ab initio TDM values of NMA upon n$\pi$* and $\pi\pi$* excitations and corresponding T-EANN predictions. It is worth noting that the incomplete T-EANN model based merely on Eq. (1) performs poorly for the n$\pi$* transition but quite well for the $\pi\pi$* transition, as evidenced by the huge RRMSE (202.60%) for n$\pi$* *versus* that (5.99%) for $\pi\pi$*. Interestingly, the disparate representability of Eq. (1) for n$\pi$* and $\pi\pi$* excitations turns out be a natural consequence of their transition characters. Indeed, the UV absorption of the NMA molecule is dominated by the peptide bond, namely H-N-C=O group. To be more specific, as illustrated in Fig. 2b, the n$\pi$* transition corresponds to a perpendicular transition from a lone pair *p* orbital of oxygen to the anti-bonding π* orbital of the N-C=O group, while the $\pi\pi$* transition takes place parallel to the N-C=O plane between the nonbonding π to the anti-bonding π* orbital. As most configurations in the data set were moderately deviated from the equilibrium geometry where the H-N-C=O group is nearly planar, the in-plane $\pi\pi$* TDM can be well described by the PDM expression in Eq. (1). In contrast, the n$\pi$* TDM vectors are almost orthogonal to the atomic coordinate vectors and largely incompatible with the implementation of Eq. (1), leading to inevitable errors. Within the proper symmetrization scheme as described in Eqs. (2-3), the T-EANN representation achieves a much better accuracy for the TDM of both n$\pi$* and $\pi\pi$* excitations, with no prior information required about the transition type. As shown in Figs. 2e-f, the RRMSE for the n$\pi$* transition is now as small as ~1.62% and the description for the $\pi\pi$* TDM is also slightly improved. This level of accuracy is much better than that of the NN model using the regular Coulomb matrix as descriptors[44] and aligning the molecule to a reference frame, especially for the n$\pi$* transition[19]. This system thus represents an excellent showcase for the importance of preserving the desired symmetry of the transition tensorial property.

We further test the quality of the T-EANN TDM model by computing the UV spectra of two proteins, namely 2bmm and 5h34, which were not included in the training set. These proteins contain so many atoms that it is prohibitively difficult to compute their spectra from atomistic molecular dynamics simulations. Instead, their spectra have been approximated by the Frenkel exciton model[45-46] in which the system is divided into many molecular excitons (*i.e.* NMA fragments) and their couplings are estimated by dipole-dipole approximation[47] (see SI for details). Impressively, as shown in Fig. 3, the T-EANN derived UV spectra of both proteins are in excellent agreement with the TDDFT counterparts, not only for the main peak positions/intensities and absorption band widths, but also for many fine structures due to exciton splitting energies that are very sensitive to the TDM vectors. Moreover, the T-EANN model is over 6 orders of magnitude faster than TDDFT. Specifically, the former takes only

~2.39 s CPU time per core for calculating both n$\pi$* and $\pi\pi$* TDMs of 10000 NMA frames, compared to $3.62 \times 10^6$ s by TDDFT calculations. These results provide convincing evidence of the high accuracy and efficiency of the T-EANN TDM model. It has been successfully applied in reproducing accurate UV spectra of realistic proteins in solutions under environmental fluctuations with excellent predictive power and transferability[48]. It should be noted that the present UV spectra have not included Franck-Condon factors from vibrational mode calculations, which will be considered in later work. We also note that full and fragmented PIP PESs have been reported for ground state trans and cis NMA[49], which, if combined with accurate TDM surfaces, could be useful for higher level calculations of electronic spectra in the future.

## IV. Conclusion

Tensorial response and transition properties of chemical systems are crucial in spectroscopic simulations and essentially determine the spectroscopic transition rules. In this work, we construct the tensorial NN model by the product of atomic coordinate vectors and virtual NN outputs (for dipole moment) or their partial derivatives with respect to atomic coordinates (for polarizability tensor). In particular, the direction of a TDM vector is taken into account by introducing a cross-product of two NN-based vectors, without the need of prior knowledge of the transition type. This treatment can be useful in representing other similar transition properties in the future. Numerical tests on PDMs, TDMs, and molecular polarizability tensors in both molecular and condensed phase systems demonstrate the high performance of this strategy, by which the corresponding vibrational and electronic spectra of large systems can be more efficiently obtained. This approach can be straightforwardly adapted to represent other important tensorial properties like magnetic dipole moment, multiple moment, and combined with any conventional atomistic NN framework.

**Acknowledgements:** This work was supported by National Key R&D Program of China (2017YFA0303500 and 2018YFA0208603), National Natural Science Foundation of China (91645202, 21722306, and 21633006), and Anhui Initiative in Quantum Information Technologies (AHY090200). We appreciate the Supercomputing Center of USTC and AM-HPC for high-performance computing services. We thank Prof Zhenggang Lan for discussion on phase correction and Prof. Yongle Li for discussion on simulations of vibrational spectra.

## References

1. Butler, K. T.; Davies, D. W.; Cartwright, H.; Isayev, O.; Walsh, A. Machine learning for molecular and materials science. *Nature* **2018,** *559* (7715), 547-555.
2. Behler, J.; Parrinello, M. Generalized neural-network representation of high-dimensional potential-energy surfaces. *Phys. Rev. Lett.* **2007,** *98*, 146401.
3. Bartók, A. P.; Payne, M. C.; Kondor, R.; Csányi, G. Gaussian approximation potentials: The accuracy of quantum mechanics, without the electrons. *Phys. Rev. Lett.* **2010,** *104* (13), 136403.
4. Bartók, A. P.; Kondor, R.; Csányi, G. On representing chemical environments. *Phys. Rev. B* **2013,** *87* (18), 184115.
5. Jiang, B.; Guo, H. Permutation invariant polynomial neural network approach to fitting potential energy surfaces. *J. Chem. Phys.* **2013,** *139*, 054112.
6. Shao, K.; Chen, J.; Zhao, Z.; Zhang, D. H. Communication: Fitting potential energy surfaces with fundamental invariant neural network. *J. Chem. Phys.* **2016,** *145* (7), 071101.
7. Schütt, K. T.; Sauceda, H. E.; Kindermans, P.-J.; Tkatchenko, A.; Müller, K.-R. SchNet – A deep learning architecture for molecules and materials. *J. Chem. Phys.* **2018,** *148* (24), 241722.
8. Zhang, L.; Han, J.; Wang, H.; Car, R.; E, W. Deep Potential Molecular Dynamics: A Scalable Model with the Accuracy of Quantum Mechanics. *Phys. Rev. Lett.* **2018,** *120* (14), 143001.
9. Zhang, Y.; Hu, C.; Jiang, B. Embedded atom neural network potentials: Efficient and accurate machine learning with a physically inspired representation. *J. Phys. Chem. Lett.* **2019,** *10* (17), 4962-4967.
10. Qu, C.; Yu, Q.; Van Hoozen, B. L.; Bowman, J. M.; Vargas-Hernández, R. A. Assessing Gaussian Process Regression and Permutationally Invariant Polynomial Approaches To Represent High-Dimensional Potential Energy Surfaces. *J. Chem. Theory Comput.* **2018,** *14* (7), 3381-3396.
11. Jiang, B.; Li, J.; Guo, H. High-Fidelity Potential Energy Surfaces for Gas-Phase and Gas–Surface Scattering Processes from Machine Learning. *J. Phys. Chem. Lett.* **2020,** *11* (13), 5120-5131.
12. Braams, B. J.; Bowman, J. M. Permutationally invariant potential energy surfaces in high dimensionality. *Int. Rev. Phys. Chem.* **2009,** *28*, 577–606.
13. Qu, C.; Yu, Q.; Bowman, J. M. Permutationally invariant potential energy surfaces. *Annu. Rev. Phys. Chem.* **2018,** *69* (1), 151-175.
14. Behler, J. Perspective: Machine learning potentials for atomistic simulations. *J. Chem. Phys.* **2016,** *145* (17), 170901.
15. Darley, M. G.; Handley, C. M.; Popelier, P. L. A. Beyond Point Charges: Dynamic Polarization from Neural Net Predicted Multipole Moments. *J. Chem. Theory Comput.* **2008,** *4* (9), 1435-1448.
16. Bereau, T.; Andrienko, D.; von Lilienfeld, O. A. Transferable Atomic Multipole Machine Learning Models for Small Organic Molecules. *J. Chem. Theory Comput.* **2015,** *11* (7), 3225-3233.
17. Liang, C.; Tocci, G.; Wilkins, D. M.; Grisafi, A.; Roke, S.; Ceriotti, M. Solvent fluctuations and nuclear quantum effects modulate the molecular hyperpolarizability of water. *Phys. Rev. B* **2017,** *96* (4), 041407.
18. Zhang, Y.; Maurer, R. J.; Guo, H.; Jiang, B. Hot-electron effects during reactive scattering of $H_2$ from Ag(111): the interplay between mode-specific electronic friction and the potential energy landscape. *Chem. Sci.* **2019,** *10* (4), 1089-1097.
19. Ye, S.; Hu, W.; Li, X.; Zhang, J.; Zhong, K.; Zhang, G.; Luo, Y.; Mukamel, S.; Jiang, J. A neural network protocol for electronic excitations of *N*-methylacetamide. *Proc. Natl. Acad. Sci. U. S. A.* **2019,**

*116* (24), 11612-11617.
20. Westermayr, J.; Gastegger, M.; Menger, M.; Mai, S.; Gonzalez, L.; Marquetand, P. Machine learning enables long time scale molecular photodynamics simulations. *Chem. Sci.* **2019,** *10* (35), 8100-8107.
21. Guan, Y.; Guo, H.; Yarkony, D. R. Extending the Representation of Multistate Coupled Potential Energy Surfaces To Include Properties Operators Using Neural Networks: Application to the 1,21A States of Ammonia. *J. Chem. Theory Comput.* **2020,** *16* (1), 302-313.
22. Spiering, P.; Meyer, J. Testing electronic friction models: Vibrational de-excitation in scattering of $H_2$ and $D_2$ from Cu(111). *J. Phys. Chem. Lett.* **2018,** *9* (7), 1803-1808.
23. Grisafi, A.; Wilkins, D. M.; Csányi, G.; Ceriotti, M. Symmetry-adapted machine learning for tensorial properties of atomistic systems. *Phys. Rev. Lett.* **2018,** *120* (3), 036002.
24. Wilkins, D. M.; Grisafi, A.; Yang, Y.; Lao, K. U.; DiStasio, R. A.; Ceriotti, M. Accurate molecular polarizabilities with coupled cluster theory and machine learning. *Proc. Natl. Acad. Sci. U. S. A.* **2019,** *116* (9), 3401-3406.
25. Willatt, M. J.; Musil, F.; Ceriotti, M. Atom-density representations for machine learning. *J. Chem. Phys.* **2019,** *150* (15), 154110.
26. Christensen, A. S.; Faber, F. A.; Lilienfeld, O. A. v. Operators in quantum machine learning: Response properties in chemical space. *J. Chem. Phys.* **2019,** *150* (6), 064105.
27. Anderson, B.; Hy, T.-S.; Kondor, R. Cormorant: Covariant Molecular Neural Networks. *arXiv e-prints*.
28. Huang, X.; Braams, B. J.; Bowman, J. M. Ab initio potential energy and dipole moment surfaces for $H_5O_2^+$. *J. Chem. Phys.* **2005,** *122*, 044308.
29. Gastegger, M.; Behler, J.; Marquetand, P. Machine learning molecular dynamics for the simulation of infrared spectra. *Chem. Sci.* **2017,** *8* (10), 6924-6935.
30. Sifain, A. E.; Lubbers, N.; Nebgen, B. T.; Smith, J. S.; Lokhov, A. Y.; Isayev, O.; Roitberg, A. E.; Barros, K.; Tretiak, S. Discovering a Transferable Charge Assignment Model Using Machine Learning. *J. Phys. Chem. Lett.* **2018,** *9* (16), 4495-4501.
31. Zhang, L.; Chen, M.; Wu, X.; Wang, H.; E, W.; Car, R. Deep neural network for the dielectric response of insulators. *arXiv e-prints*.
32. Unke, O. T.; Meuwly, M. PhysNet: A Neural Network for Predicting Energies, Forces, Dipole Moments, and Partial Charges. *J. Chem. Theory Comput.* **2019,** *15* (6), 3678-3693.
33. Medders, G. R.; Paesani, F. Infrared and Raman Spectroscopy of Liquid Water through "First-Principles" Many-Body Molecular Dynamics. *J. Chem. Theory Comput.* **2015,** *11* (3), 1145-1154.
34. Wang, Y.; Huang, X.; Shepler, B. C.; Braams, B. J.; Bowman, J. M. Flexible, ab initio potential, and dipole moment surfaces for water. I. Tests and applications for clusters up to the 22-mer. *J. Chem. Phys.* **2011,** *134*, 094509.
35. Zuo, Y.; Chen, C.; Li, X.; Deng, Z.; Chen, Y.; Behler, J.; Csányi, G.; Shapeev, A. V.; Thompson, A. P.; Wood, M. A.; Ong, S. P. Performance and Cost Assessment of Machine Learning Interatomic Potentials. *J. Phys. Chem. A* **2020,** *124* (4), 731-745.
36. Cheng, B.; Engel, E. A.; Behler, J.; Dellago, C.; Ceriotti, M. Ab initio thermodynamics of liquid and solid water. *Proceedings of the National Academy of Sciences* **2019,** *116* (4), 1110-1115.
37. Zhang, Y.; Hu, C.; Jiang, B. Bridging the Efficiency Gap Between Machine Learned Ab Initio Potentials and Classical Force Fields. *arXiv e-prints* **2020**, arXiv:2006.16482.
38. Zhang, Y.; Maurer, R. J.; Jiang, B. Symmetry-Adapted High Dimensional Neural Network

Representation of Electronic Friction Tensor of Adsorbates on Metals. *J. Phys. Chem. C* **2020,** *124* (1), 186-195.
39. Westermayr, J.; Gastegger, M.; Marquetand, P. Combining SchNet and SHARC: The SchNarc Machine Learning Approach for Excited-State Dynamics. *J. Phys. Chem. Lett.* **2020,** *11* (10), 3828-3834.
40. Sommers, G. M.; Calegari Andrade, M. F.; Zhang, L.; Wang, H.; Car, R. Raman spectrum and polarizability of liquid water from deep neural networks. *Phys. Chem. Chem. Phys.* **2020,** *22* (19), 10592-10602.
41. Wang, L.; Middleton, C. T.; Zanni, M. T.; Skinner, J. L. Development and Validation of Transferable Amide I Vibrational Frequency Maps for Peptides. *J. Phys. Chem. B* **2011,** *115* (13), 3713-3724.
42. Frisch, M. J.; Trucks, G. W.; Schlegel, H. B.; Scuseria, G. E.; Robb, M. A.; Cheeseman, J. R.; Scalmani, G.; Barone, V.; Petersson, G. A.; Nakatsuji, H.; Li, X.; Caricato, M.; Marenich, A. V.; Bloino, J.; Janesko, B. G.; Gomperts, R.; Mennucci, B.; Hratchian, H. P.; Ortiz, J. V.; Izmaylov, A. F.; Sonnenberg, J. L.; Williams; Ding, F.; Lipparini, F.; Egidi, F.; Goings, J.; Peng, B.; Petrone, A.; Henderson, T.; Ranasinghe, D.; Zakrzewski, V. G.; Gao, J.; Rega, N.; Zheng, G.; Liang, W.; Hada, M.; Ehara, M.; Toyota, K.; Fukuda, R.; Hasegawa, J.; Ishida, M.; Nakajima, T.; Honda, Y.; Kitao, O.; Nakai, H.; Vreven, T.; Throssell, K.; Montgomery Jr., J. A.; Peralta, J. E.; Ogliaro, F.; Bearpark, M. J.; Heyd, J. J.; Brothers, E. N.; Kudin, K. N.; Staroverov, V. N.; Keith, T. A.; Kobayashi, R.; Normand, J.; Raghavachari, K.; Rendell, A. P.; Burant, J. C.; Iyengar, S. S.; Tomasi, J.; Cossi, M.; Millam, J. M.; Klene, M.; Adamo, C.; Cammi, R.; Ochterski, J. W.; Martin, R. L.; Morokuma, K.; Farkas, O.; Foresman, J. B.; Fox, D. J. *Gaussian 16 Rev. B.01*, Wallingford, CT, 2016.
43. Berman, H. M.; Westbrook, J.; Feng, Z.; Gilliland, G.; Bhat, T. N.; Weissig, H.; Shindyalov, I. N.; Bourne, P. E. The Protein Data Bank. *Nucleic Acids Res.* **2000,** *28* (1), 235-242.
44. Rupp, M.; Tkatchenko, A.; Müller, K.-R.; von Lilienfeld, O. A. Fast and accurate modeling of molecular atomization energies with machine learning. *Phys. Rev. Lett.* **2012,** *108* (5), 058301.
45. Abramavicius, D.; Palmieri, B.; Mukamel, S. Extracting single and two-exciton couplings in photosynthetic complexes by coherent two-dimensional electronic spectra. *Chem. Phys.* **2009,** *357* (1), 79-84.
46. Abramavicius, D.; Jiang, J.; Bulheller, B. M.; Hirst, J. D.; Mukamel, S. Simulation Study of Chiral Two-Dimensional Ultraviolet Spectroscopy of the Protein Backbone. *J. Am. Chem. Soc.* **2010,** *132* (22), 7769-7775.
47. Kasha, M.; Rawls, H. R.; El-Bayoumi, M. A. The exciton model in molecular spectroscopy. *Pure Appl. Chem.* **1965,** *11* (3-4), 371-392.
48. Jinxiao Zhang; Sheng Ye; Kai Zhong; Yaolong Zhang; Yuanyuan Chong; Luyuan Zhao; Huiting Zhou; Sibei Guo; Guozhen Zhang; Bin Jiang; Shaul Mukamel; Jiang, J. A Machine-Learning Protocol for Ultraviolet Protein-backbone Spectroscopy under Environmental Fluctuations. *Chem. Sci.* **submitted**.
49. Nandi, A.; Qu, C.; Bowman, J. M. Full and fragmented permutationally invariant polynomial potential energy surfaces for trans and cis N-methyl acetamide and isomerization saddle points. *J. Chem. Phys.* **2019,** *151* (8), 084306.

Table I. Comparison of RRMSEs for permanent dipole moment and polarizability in several water-related systems over randomly chosen 500 configurations in the test set, with the T-EANN and SA-GAP[23] models. The numbers represent average errors of randomly chosen 500 configurations in the test set.

| System | $\vec{\mu}$ | | $\alpha$ | |
|---|---|---|---|---|
| | T-EANN | SA-GAP* | T-EANN | SA-GAP* |
| $H_2O$ | 0.02% | ~0.11% | 0.02% | ~0.02%/0.12% |
| $(H_2O)_2$ | 6.6% | ~5.3% | 4.2% | ~6.4%/7.8% |
| $(H_5O_2)^+$ | 1.3% | ~2.4% | 0.3% | ~3.8%/0.97% |
| Liquid water | 16% | \ | 2.2% | ~5.8%/19%** |

* The SA-GAP method[23] is based on spherical tensors in its irreducible spherical tensor (IST) representation with the covariant λ-SOAP kernels. There are three IST components for dipole moment (λ=1) and six IST components for polarizability tensor (one for λ=0 and five for λ=2, leading to respective RRMSEs in this Table).

**RRMSEs were reported for dielectric response tensors by indirect learning of molecular polarizability[23], which are used here for qualitative comparison only.

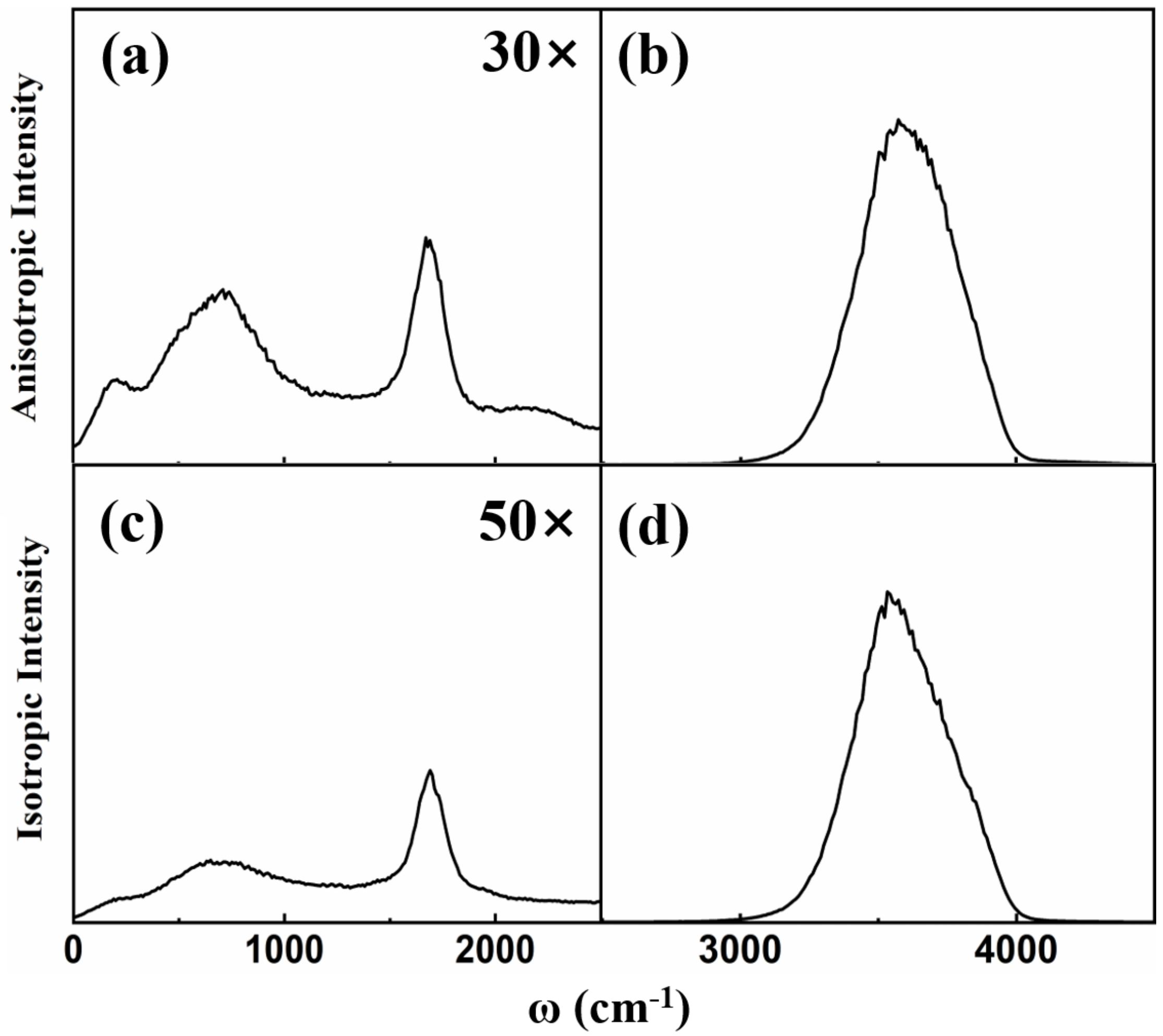

Fig. 1 (a), (b) Anisotropic and (c), (d) isotropic reduced Raman intensities of liquid water at *T*=300K obtained from classical molecular dynamics with the EANN potential at revPBE0-D3 level and the T-EANN polarizability model at PBE level. To better visualize the line shape in low frequency regime, spectral intensities are magnified by a factor of 30 (anisotropic) and 50 (isotropic), respectively.

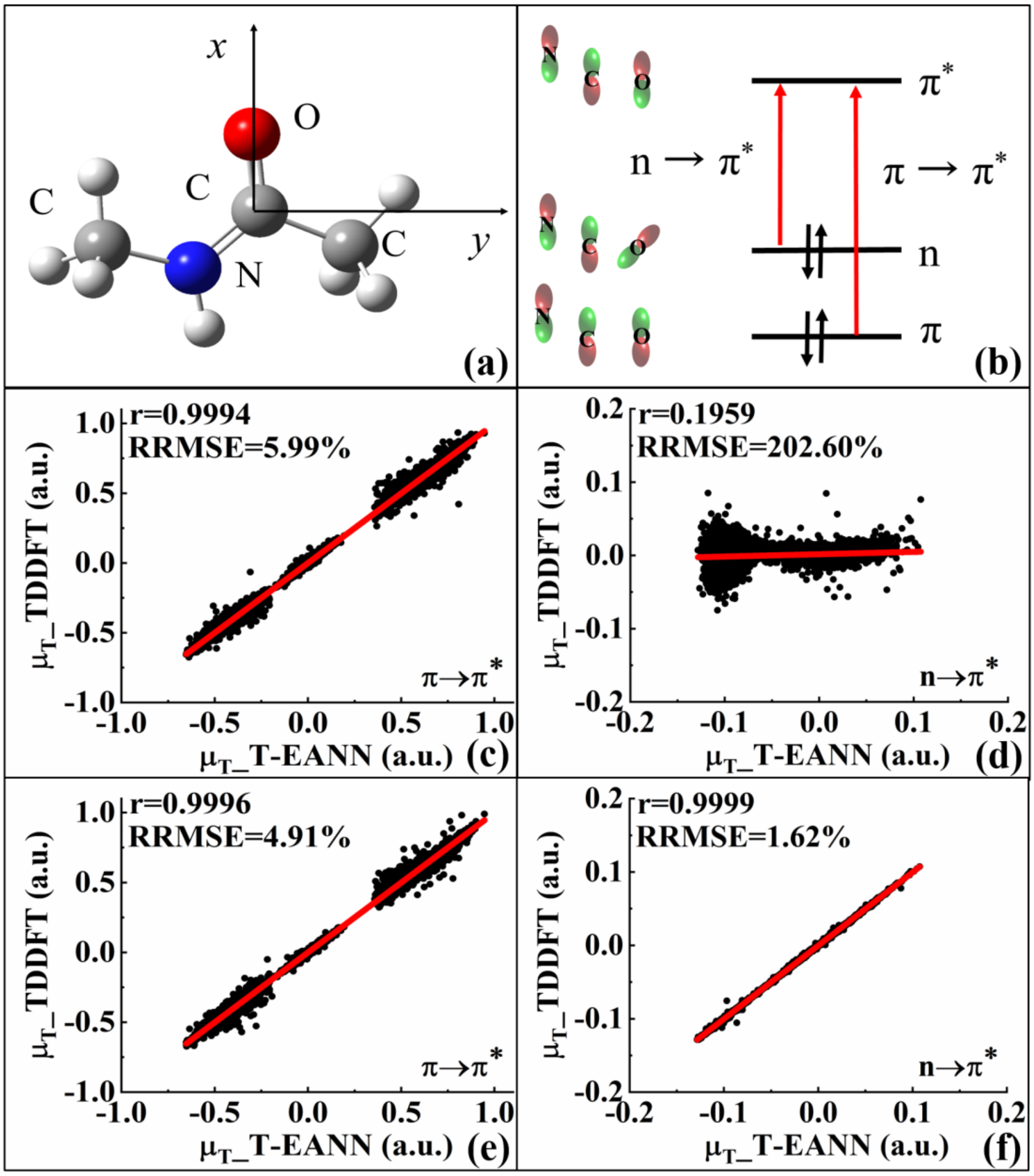

Fig. 2 (a) Molecular geometry of NMA with the peptide bond placed in the *xy* plane and (b) schematic diagram of its $\pi \rightarrow \pi^*$ and $n \rightarrow \pi^*$ electronic transitions. (c-f) Comparison of the TDDFT results of $\pi \rightarrow \pi^*$ and $n \rightarrow \pi^*$ transition dipole moments ($\vec{\boldsymbol{\mu}}_T$) versus the T-EANN predictions obtained from the incomplete model based on Eq. (1) only (c-d) and the correct one based on Eqs. (2-3) (e-f), respectively.

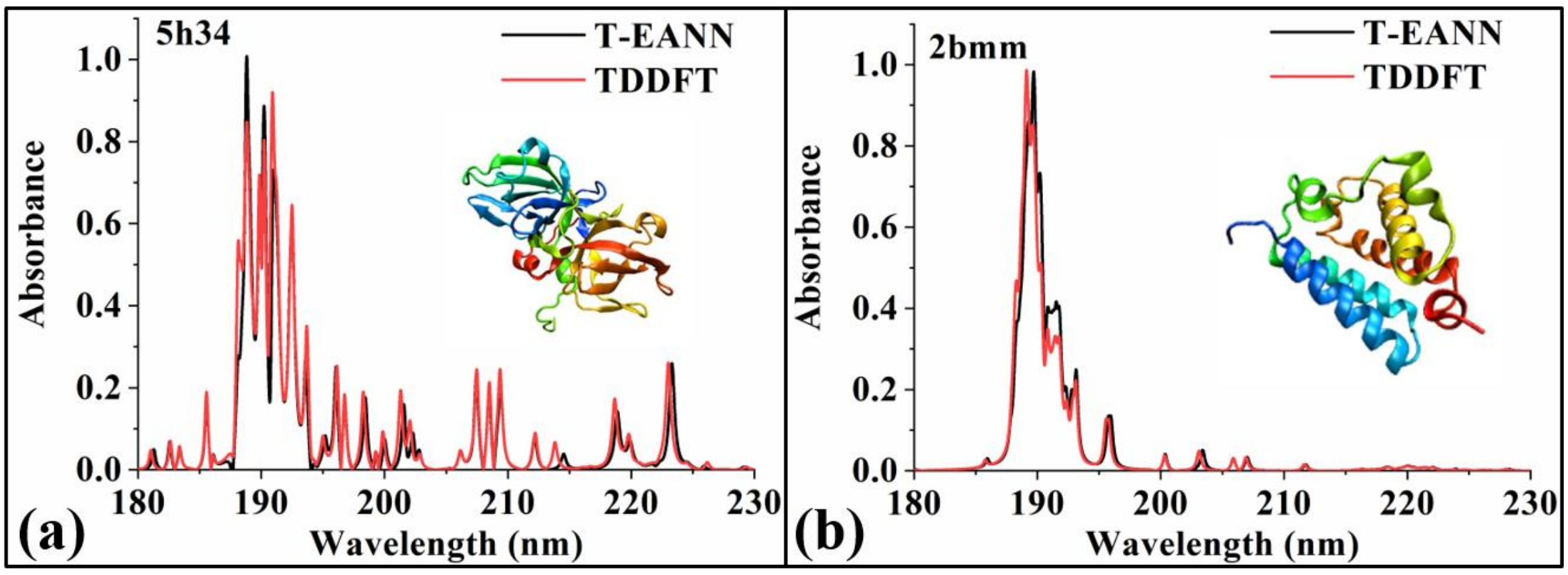

Fig. 3 UV adsorption spectra of proteins (a) 5h34 and (b) 2bmm calculated by the T-EANN model (black) and TDDFT (red).

**Supporting Information**

**for**

# Efficient and Accurate Spectroscopic Simulations with Symmetry-Preserving Neural Network Models for Tensorial Properties

Yaolong Zhang[1,2], Sheng Ye[1,3], Jinxiao Zhang[1,3], Ce Hu[1,2], Jun Jiang[1,3], Bin Jiang[1,2*]

[1]*Hefei National Laboratory for Physical Science at the Microscale, Department of Chemical Physics, University of Science and Technology of China, Hefei, Anhui 230026, China*

[2]*Key Laboratory of Surface and Interface Chemistry and Energy Catalysis of Anhui Higher Education Institutes, University of Science and Technology of China, Hefei, Anhui 230026, China*

[3]*Chinese Academy of Sciences Center for Excellence in Nanoscience, University of Science and Technology of China, Hefei, Anhui 230026, China*

**Supplementary Methods**

## A. Generalized embedded atom neural network representation

As discussed in the main text, any atomistic NN framework for scalar quantities can be readily adapted to represent tensorial properties with our proposed algorithm and one needs no modification of the symmetry-invariant descriptors. In this work, we employed our recently proposed embedded atom neural network (EANN) model[1]. The EANN model is inspired from the physically-derived embedded atom method (EAM)[2] force field. The total energy of a system is the sum of the atomic energy, each of which is a function of electron density at the central atom position embedded in the surrounding environment. Atomic NNs thus represent the complex relationship between the embedded density and the atomic energy,

$$E = \sum_{i=1}^{N} E_i^{NN}(\boldsymbol{\rho}^i), \tag{S1}$$

where $N$ is the total number of atoms in the system and $\boldsymbol{\rho}^i$ is electron density vector at $i$th atom, which can be approximated by the square of the linear combination of different atomic orbitals,

$$\rho_{L,\alpha,r_s}^{i} = \sum_{l_x,l_y,l_z}^{l_x+l_y+l_z=L} \frac{L!}{l_x!l_y!l_z!} \left(\sum_{j=1}^{n_{\mathrm{atom}}} c_j \varphi_{l_x l_y l_z}^{\alpha,r_s}(\mathbf{r}^{ij})\right)^2 . \tag{S2}$$

where $c_j$ is the orbital coefficient of atom $j$ and is optimized in the training process. $n_{\mathrm{atom}}$ is the number of neighboring atoms. $\varphi_{l_x l_y l_z}^{\alpha,r_s}(\mathbf{r}^{ij})$ is Gaussian-type like (GTO) orbitals

$$\varphi_{l_x l_y l_z}^{\alpha,r_s}(\mathbf{r}) = x^{l_x} y^{l_y} z^{l_z} exp(-\alpha|r - r_s|^2), \tag{S3}$$

where $\mathbf{r}$=($x$, $y$, $z$) is the coordinates vector of an electron relative to the corresponding center atom, $r$ is the norm of the vector, $\alpha$ and $r_s$ are parameters used for tuning radial distributions of atomic orbitals, $L$ represents the orbital angular momentum, which is the summation of angular moment along each axis $l_x$, $l_y$, $l_z$. In practice, a cutoff function $f_c(r_{ij})$ is multiplied to each GTO to ensure that the contribution of each neighbor atom decays smoothly to zero at $r_c$,[3]

$$f_c(r)=\begin{cases}[0.5+0.5\cos(\pi r/r_c)]^2, & r\le r_c\\ 0, & r>r_c\end{cases}. \tag{S4}$$

These hyper-parameters used to define the density-like descriptors for each system in this work are listed in Table S1 ~ S5.

By construction, these embedded density-like descriptors can preserve the translational, rotational and permutational symmetry of potential energy within the complete nuclear permutation and inversion (CNPI) group in an efficient way[4]. Through a straightforward transformation[4], these descriptors include the angular information with a two-body computational cost in comparison to the commonly used atom centered symmetry functions[5]. We thus need to consider only the covariance of the tensorial EANN (T-EANN) models with respect to rotation for permanent dipole moment (PDM), transition dipole moment (TDM), as well as molecular polarizability, that is subject to the three dimensional rotation group SO(3). Specifically, these physical quantities transform under a rotation (represented by a transformation matrix $\mathbf{R}$) as,

$$\vec{\boldsymbol{\mu}} \mapsto \mathbf{R}\vec{\boldsymbol{\mu}},\ \vec{\boldsymbol{\mu}}_{\mathrm{T}} \mapsto \mathbf{R}\vec{\boldsymbol{\mu}}_{\mathrm{T}},\ \boldsymbol{\alpha} \mapsto \mathbf{R}\boldsymbol{\alpha}\mathbf{R}^{\mathrm{T}}, \tag{S5}$$

where $\mathbf{R}^{\mathrm{T}}$ is the transpose of $\mathbf{R}$. As we stated in the main text, our approach takes no covariant basis (or kernel) functions under SO(3) group operations[6], which would be scrambled by the nonlinearity of NNs anyways. Instead, we simply take advantage of the Cartesian coordinate vector or the derivative of a scalar property with respect to it (behaving like the force vector), which is naturally compatible with SO(3) symmetry when the molecule is rotated. In other words, by multiplying the scalar NN output with or differentiating the scalar NN output with respect to the Cartesian coordinate vector, the resultant NN-based tensors have the same transformation rules, *e.g.*,

$$q_i\vec{\mathbf{r}}_i \mapsto \mathbf{R}q_i\vec{\mathbf{r}}_i, \ \frac{\partial q_i}{\partial \vec{\mathbf{r}}_i} \mapsto \mathbf{R}\frac{\partial q_i}{\partial \vec{\mathbf{r}}_i}, \quad \mathbf{D}(\mathbf{D}^{\mathrm{T}}) \mapsto \mathbf{R}\mathbf{D}(\mathbf{D}^{\mathrm{T}})\mathbf{R}^{\mathrm{T}}, \tag{S6}$$

where we take $\boldsymbol{\alpha}^{\mathrm{NN1}}$ in Eq. (7) only for illustrating the tensorial representation of polarizability. As a result, our tensorial EANN models fully capture the symmetry of these molecular properties. As discussed below, one would not get a faithful fit using symmetry deficient models, while just relying on alleged virtues of the training itself. We note that our approach is a systematic extension of previous machine learning models for PDM[7-12] and is similar to the deep NN model for polarizability[13].

The EANN model for a scalar property can be trained by minimizing the following cost function,

$$S(\mathbf{w}) = \sum_{m=1}^{N_{data}} \left| E_m^{NN} - E_m^{QC} \right|^2 / N_{data}. \tag{S7}$$

where $\mathbf{w}$ is the collection of the NN weight parameters and $N_{\mathrm{data}}$ is the size of training dataset, $E_m^{NN}$ and $E_m^{QC}$ are the sum of atomic NN energies of the *mth* data point and the corresponding target total energy obtained from quantum chemical (QC) calculations. In T-EANN models, since the geometric descriptors remain unchanged, one just needs to slightly modify the objective function to be minimized. As a result, the T-EANN models for dipole moments can be trained in the following way,

$$S(\mathbf{w}) = \sum_{m=1}^{N_{data}} \left| \vec{\boldsymbol{\mu}}_m^{NN} - \vec{\boldsymbol{\mu}}_m^{QC} \right|^2 / N_{data}. \tag{S8}$$

where $\vec{\boldsymbol{\mu}}_m^{NN}$ is the sum of all atomic dipole moments via Eq. (1) for PDM or Eqs (2-3)

for TDM, and $\vec{\boldsymbol{\mu}}_m^{QC}$ is corresponding QC-calculated PDM or TDM vector. For the polarizability tensor, similarly, we minimize the objective function expressed as

$$S(\mathbf{w}) = \sum_{m=1}^{N_{data}} \left|\boldsymbol{\alpha}_m^{NN} - \boldsymbol{\alpha}_m^{QC}\right|^2 / N_{data} . \tag{S9}$$

More details about the EANN model can be found in our previous work[1].

To quantify the quality of the NN fit, given the wide numerical ranges of the tensorial properties, we present in this work both the absolute root-mean-squared-error (RMSE) and the RMSE relative to intrinsic standard deviation ($\sigma$) of the testing samples (RRMSE)[6]. Specifically,

$$\text{RMSE}=\sqrt{\frac{1}{N_{data}} \sum_{m=1}^{N_{data}} \left|\boldsymbol{\Gamma}^{NN} - \boldsymbol{\Gamma}^{QC}\right|^2} , \tag{S10}$$

$$\text{RRMSE}=\frac{\sqrt{\frac{1}{N_{data}} \sum_{m=1}^{N_{data}} \left|\boldsymbol{\Gamma}^{NN} - \boldsymbol{\Gamma}^{QC}\right|^2}}{\sigma(\boldsymbol{\Gamma})} , \tag{S11}$$

and the standard deviation is given by,

$$\sigma(\boldsymbol{\Gamma})=\sqrt{\frac{1}{N_{data}-1} \sum_{m=1}^{N_{data}} \left|\boldsymbol{\Gamma}^{QC} - \left\langle \boldsymbol{\Gamma}^{QC} \right\rangle\right|^2} . \tag{S12}$$

Here $\boldsymbol{\Gamma}$ represent a tensorial property, namely $\vec{\boldsymbol{\mu}}$, $\vec{\boldsymbol{\mu}}_{\mathrm{T}}$ or $\boldsymbol{\alpha}$ in this work.

## B. Electronic structure calculations

The permanent dipole moment (**μ**) and polarizability (**α**) data of $H_2O$, $(H_2O)_2$, $H_5O_2^+$ and liquid water have been calculated by Ceriotti and coworkers[6]. We refer the readers to Ref. 6 for more computational details. Briefly, ab initial calculations for $H_2O$, $(H_2O)_2$, and $H_5O_2^+$ were performed at the CCSD/d-aug-cc-pvtz level using Dalton 2016[14]. Liquid water was described by a 32-water box and calculated at the DFT/PBE-USPP level, using a plane-wave basis set with a kinetic energy cutoff of 55 Ry. The first

Brillouin zone of the periodic system has been sampled with 5 k-points along each reciprocal lattice direction[6]. Overall, 1000 configurations were collected for each system and half of them were randomly chosen for training, leaving the rest 500 points as test set.

The transition dipole moment data for nπ* and ππ* transitions of N-methylacetamide (NMA) were calculated by time-dependent density functional theory (TDDFT) implemented in the Gaussian 16 package[15], at the PBE0/cc-pVDZ level. We extracted 50000 NMA configurations from 1000 different protein backbones at room temperature from the RCSB Protein Data Bank (PDB)[16]. Phase corrections have been carefully done by evaluating the wavefunction overlaps of neighboring configurations along the trajectories[17].

### C. Frenkel exciton model

Proteins consist of peptides and amino acid residues. It is known that ultraviolet absorption (UV) spectra of proteins are largely due to the electronic excitation of peptides in protein backbones. In this work, we split the protein into different peptide fragments and amino acid residues. Frenkel exciton model[18-19] is employed to construct the exciton Hamiltonian derived from the interaction between different peptides and amino acid residues,

$$\hat{H} = \sum_{ma} \varepsilon_{ma} \hat{B}_{ma}^{\dagger} \hat{B}_{ma} + \sum_{ma,nb}^{m \neq n} J_{ma,nb} \hat{B}_{ma}^{\dagger} \hat{B}_{nb} , \tag{S13}$$

where $m$ and $n$ represent different peptide fragments, $a$ and $b$ correspond to different transition types (*i.e.*, nπ* (perpendicular) and ππ* (parallel) transitions in this work), $\hat{B}_{ma}^{\dagger}$ and $\hat{B}_{ma}$ are the electron creation and annihilation operators of corresponding transitions, $\varepsilon_{ma}$ is the excitation energy of the *mth* peptide coupled with neighboring amino acid residues, and $J_{ma,nb}$ is the coupling between excited states of two peptide fragments. More details about this model can be found elsewhere[18-19].

In principle, an accurate depiction of these couplings require very time-consuming two-electron integral and excited state calculations. Dipole-dipole approximation is regarded as a common and simple way to estimate the interaction between electrons[20]. To this end, $\varepsilon_{ma}$ and $J_{ma,nb}$ can be approximated as,

$$\varepsilon_{ma} = \varepsilon_{ma}^{0} + \sum_{k} \frac{1}{4\pi\varepsilon\varepsilon_0} \left( \frac{\vec{\boldsymbol{\mu}}_{\mathrm{T},ma} \cdot \vec{\boldsymbol{\mu}}^{k}}{\left|\mathbf{r}_{mk}\right|^3} - 3\frac{(\vec{\boldsymbol{\mu}}_{\mathrm{T},ma} \cdot \mathbf{r}_{mk})(\vec{\boldsymbol{\mu}}^{k} \cdot \mathbf{r}_{mk})}{\left|\mathbf{r}_{mk}\right|^5} \right), \tag{S14}$$

$$J_{ma,nb} = \sum_{m,n}^{m \neq n} \frac{1}{4\pi\varepsilon\varepsilon_0} \left( \frac{\vec{\boldsymbol{\mu}}_{\mathrm{T},ma} \cdot \vec{\boldsymbol{\mu}}_{\mathrm{T},nb}}{\left|\mathbf{r}_{mn}\right|^3} - 3\frac{(\vec{\boldsymbol{\mu}}_{\mathrm{T},ma} \cdot \mathbf{r}_{mn})(\vec{\boldsymbol{\mu}}_{\mathrm{T},nb} \cdot \mathbf{r}_{mn})}{\left|\mathbf{r}_{mn}\right|^5} \right). \tag{S15}$$

Here, $\varepsilon_{ma}^{0}$ is the excitation energy of an isolated peptide (*i.e.*, the NMA molecule in this work), $\vec{\boldsymbol{\mu}}_{\mathrm{T},ma}$ and $\vec{\boldsymbol{\mu}}_{\mathrm{T},nb}$ are the TDMs of the corresponding peptide, $\vec{\boldsymbol{\mu}}^{k}$ is the PDM of an amino acid residue, $\mathbf{r}_{mk} / \mathbf{r}_{mn}$ is the position vector of a peptide referred to an amino acid residue or another peptide. To further simplify the simulations, we have ignored the coupling between peptides and amino acid residues because of their negligible influence[18]. In comparison of ab initio based and NN predicted UV spectra of two real proteins, namely 2bmm and 5h34, $\varepsilon_{ma}^{0}$, $\vec{\boldsymbol{\mu}}_{\mathrm{T},ma}$, and $\vec{\boldsymbol{\mu}}_{\mathrm{T},nb}$ generated from TDDFT calculations are represented by scalar and tensorial EANN models, respectively. The scalar EANN model of $\varepsilon_{ma}^{0}$ is essentially the same as that for fitting potential energy surfaces, yielding an average RMSE of 7.8 meV and 5.1 meV for nπ* and ππ* transitions, which are sufficiently accurate for calculating the UV spectra. The accuracy of the T-EANN model for TDM has been presented in the main text.

## D. Molecular dynamics simulations of Raman spectra for liquid water

To obtain the Raman spectra of liquid water, we have performed classical molecular dynamics (MD) simulations at T=300K in a cubic box with its side length of 12.42 Å including 64 water molecules. The system was first equilibrated within 5 ps by the Andersen thermostat[21], followed by 200 ps NVE MD simulations with a time step of 0.05 fs. It is well-known that the differential cross section of Raman scattering is related to the Fourier transform of the autocorrelation function of polarizability tensor[22]. In experiments, the reduced Raman intensity $R(\omega)$ is typically reported, which includes the isotropic and anisotropic components that can be both obtained from MD simulations as follows,

$$R_{iso}(\omega) \propto \omega \tanh(\hbar\omega / 2kT) \int_{-\infty}^{+\infty} e^{-i\omega t} \langle \alpha_{iso}(0)\alpha_{iso}(t) \rangle dt , \tag{S16}$$

and,

$$R_{aniso}(\omega) \propto \omega \tanh(\hbar\omega / 2kT) \int_{-\infty}^{+\infty} e^{-i\omega t} \mathrm{Tr} \langle \boldsymbol{\beta}(0)\boldsymbol{\beta}(t) \rangle dt , \tag{S17}$$

where $\alpha_{iso}$ and $\boldsymbol{\beta}$ are isotropic and anisotropic components of the polarizability tensor, with $\alpha_{iso} \equiv \mathbf{Tr}(\boldsymbol{\alpha})/3$, $\boldsymbol{\beta}=\boldsymbol{\alpha}-\alpha_{iso}\mathbf{I}$, the brackets are replaced by the time average of the product of corresponding polarizability tensors, and the integrals are replaced by the discrete Fourier transform over the real time domain. In practice, the Fourier transform of the polarizability autocorrelation is more frequently replaced by the Fourier transform of the autocorrelation of the polarizability time derivative[23], namely,

$$\int_{-\infty}^{+\infty} e^{-i\omega t} \langle \dot{\alpha}_{iso}(0)\dot{\alpha}_{iso}(t) \rangle dt = \omega^2 \int_{-\infty}^{+\infty} e^{-i\omega t} \langle \alpha_{iso}(0)\alpha_{iso}(t) \rangle dt , \tag{S18}$$

and,

$$\int_{-\infty}^{+\infty} e^{-i\omega t} \mathrm{Tr} \langle \dot{\boldsymbol{\beta}}(0)\dot{\boldsymbol{\beta}}(t) \rangle dt = \omega^2 \int_{-\infty}^{+\infty} e^{-i\omega t} \mathrm{Tr} \langle \boldsymbol{\beta}(0)\boldsymbol{\beta}(t) \rangle dt , \tag{S19}$$

We have run ten independent sets of trajectory simulations whose average give the Raman spectra shown in Fig. 1. A Hann window function with 300 fs width was imposed to smooth the spectra.

Our EANN potential for liquid water was constructed using the ab initio data calculated by the Cheng *et al.* at revPBE0-D3 level, including energies and forces of 1593 configurations[24]. Our EANN potential gives a test RMSE 2.0 meV/atom for energy and 129meV/ Å for atomic forces comparable to the accuracy of the NN potential reported in that work[24]. More details about the EANN water potential can be found in our recent preprint[25].

## Supplementary Results and Discussion

### A. Numerical verification of the T-EANN model for polarizability tensor

In the T-EANN representation of $\boldsymbol{\alpha}$, we need three terms $\boldsymbol{\alpha}^{\mathrm{NN1}}$, $\boldsymbol{\alpha}^{\mathrm{NN2}}$ and $\boldsymbol{\alpha}^{\mathrm{NN3}}$ to guarantee the proper property of an arbitrary polarizability tensor. Taking water monomer as an illustrative example, we demonstrate numerically how each of these terms contributes. The molecule is placed in the *x*-*z* plane and its molecular geometry can be expressed in terms of atomic Cartesian coordinate vectors as,

$$\mathbf{r}=\begin{bmatrix} x_{\mathrm{O}} & y_{\mathrm{O}} & z_{\mathrm{O}} \\ x_{\mathrm{H1}} & y_{\mathrm{H1}} & z_{\mathrm{H1}} \\ x_{\mathrm{H2}} & y_{\mathrm{H2}} & z_{\mathrm{H2}} \end{bmatrix}. \tag{S20}$$

We take one arbitrary configuration from the training data set,

$$\mathbf{r}=\begin{bmatrix} 0.057536 & 0.000000 & -0.033219 \\ -0.841234 & -0.000000 & -0.402656 \\ -0.071907 & 0.000000 & 0.929858 \end{bmatrix}. \tag{S21}$$

and its ab initio polarizability tensor reads,

$$\boldsymbol{\alpha}^{QC}=\begin{bmatrix}9.931967 & 0.000000 & 0.205256\\ 0.000000 & 9.459317 & -0.000000\\ 0.205256 & -0.000000 & 10.168985\end{bmatrix}. \tag{S22}$$

As clearly seen, $\boldsymbol{\alpha}^{QC}$ is a full-rank symmetric matrix although $H_2O$ is completely planar. However, by construction, $\boldsymbol{\alpha}^{\mathrm{NN1}}$ and $\boldsymbol{\alpha}^{\mathrm{NN2}}$ are both rank-deficient matrices. After successful training process, the T-EANN model contains the following $\boldsymbol{\alpha}^{\mathrm{NN1}}$ and $\boldsymbol{\alpha}^{\mathrm{NN2}}$ matrices,

$$\boldsymbol{\alpha}^{\mathrm{NN1}}=\begin{bmatrix}0.744341 & 0.000000 & 0.024220\\ 0.000000 & 0.000000 & 0.000000\\ 0.024220 & 0.000000 & 0.772309\end{bmatrix}, \tag{S23}$$

$$\boldsymbol{\alpha}^{\mathrm{NN2}}=\begin{bmatrix}-0.271611 & 0.000000 & 0.181350\\ 0.000000 & 0.000000 & 0.000000\\ 0.181350 & 0.000000 & -0.062198\end{bmatrix}, \tag{S24}$$

both with zero value for $\alpha_{yy}$. However, the scalar matrix $\boldsymbol{\alpha}^{\mathrm{NN3}}$ adequately solves this issue,

$$\boldsymbol{\alpha}^{\mathrm{NN3}}=\begin{bmatrix}9.459202 & 0.000000 & 0.000000\\ 0.000000 & 9.459202 & 0.000000\\ 0.000000 & 0.000000 & 9.459202\end{bmatrix}. \tag{S25}$$

The overall T-EANN polarizability tensor is the sum of these three terms,

$$\boldsymbol{\alpha}^{\mathrm{NN}}=\begin{bmatrix}9.931932 & 0.000000 & 0.205570\\ 0.000000 & 9.459202 & 0.000000\\ 0.205570 & 0.000000 & 10.169314\end{bmatrix}, \tag{S26}$$

which is in excellent agreement with the ab initio value $\boldsymbol{\alpha}^{QC}$.

We further compare the performance of incomplete T-EANN models, *i.e.* fitting the ab initio $\boldsymbol{\alpha}$ values for $H_2O$, $(H_2O)_2$, and $(H_5O_2)^+$ molecules with $\boldsymbol{\alpha}^{\mathrm{NN1}}$, $\boldsymbol{\alpha}^{\mathrm{NN1}}+\boldsymbol{\alpha}^{\mathrm{NN2}}$, $\boldsymbol{\alpha}^{\mathrm{NN1}}+\boldsymbol{\alpha}^{\mathrm{NN3}}$, and the full T-EANN model, respectively. As seen in Table S6, using $\boldsymbol{\alpha}^{\mathrm{NN1}}$ only results in huge prediction errors, especially for water and water dimer. This is understandable as they contain either completely or mostly planar configurations. Similar results are observed when using $\boldsymbol{\alpha}^{\mathrm{NN2}}$ only (not shown). An accompanying issue is that "over-fitting" severely takes place when training this single term. Including $\boldsymbol{\alpha}^{\mathrm{NN2}}$ leads to a marginal improvement but largely remedies the over-fitting issue, making the training process more stable. Interestingly, adding $\boldsymbol{\alpha}^{\mathrm{NN3}}$ substantially lowers the prediction errors, suggesting that the singularity of $\boldsymbol{\alpha}^{\mathrm{NN1}}$ and $\boldsymbol{\alpha}^{\mathrm{NN2}}$ at planar configurations is problematic. Incorporating all three terms fix this problem and provide sufficient repeatability of the T-EANN model, which further considerably improves the results. This problem still exists for a more complex system $H_5O_2^+$, though less severe because fewer configurations are planar. Also compared in Table S6 are the results of using $\boldsymbol{\alpha}^{\mathrm{NN1}}$ based on the derivative matrix or $\boldsymbol{\alpha}^{\mathrm{NN1'}}$ based on the dipole-like vector. It is found that the former performs somewhat better than latter, especially for $H_5O_2^+$. The better performance of the derivative matrix is due possibly to that it follows more closely as the physical definition of induced dipole, *i.e.* the derivative of potential energy with respect to electric field vector. We note that transforming the tensor from the Cartesian coordinate space to the irreducible spherical representation[26] could eliminate the singularity problem in planar configurations.[13]

## Reference

1. Zhang, Y.; Hu, C.; Jiang, B. Embedded atom neural network potentials: Efficient and accurate machine learning with a physically inspired representation. *J. Phys. Chem. Lett.* **2019,** *10* (17), 4962-4967.
2. Daw, M. S.; Baskes, M. I. Embedded-atom method: Derivation and application to impurities, surfaces, and other defects in metals. *Phys. Rev. B* **1984,** *29* (12), 6443-6453.
3. Behler, J. Atom-centered symmetry functions for constructing high-dimensional neural network potentials. *J. Chem. Phys.* **2011,** *134*, 074106.
4. Takahashi, A.; Seko, A.; Tanaka, I. Conceptual and practical bases for the high accuracy of machine learning interatomic potentials: Application to elemental titanium. *Phys. Rev. Materials* **2017,** *1* (6), 063801.
5. Behler, J.; Parrinello, M. Generalized neural-network representation of high-dimensional potential-energy surfaces. *Phys. Rev. Lett.* **2007,** *98*, 146401.
6. Grisafi, A.; Wilkins, D. M.; Csányi, G.; Ceriotti, M. Symmetry-adapted machine learning for tensorial properties of atomistic systems. *Phys. Rev. Lett.* **2018,** *120* (3), 036002.
7. Huang, X.; Braams, B. J.; Bowman, J. M. Ab initio potential energy and dipole moment surfaces for $H_5O_2^+$. *J. Chem. Phys.* **2005,** *122*, 044308.
8. Gastegger, M.; Behler, J.; Marquetand, P. Machine learning molecular dynamics for the simulation of infrared spectra. *Chem. Sci.* **2017,** *8* (10), 6924-6935.
9. Sifain, A. E.; Lubbers, N.; Nebgen, B. T.; Smith, J. S.; Lokhov, A. Y.; Isayev, O.; Roitberg, A. E.; Barros, K.; Tretiak, S. Discovering a Transferable Charge Assignment Model Using Machine Learning. *J. Phys. Chem. Lett.* **2018,** *9* (16), 4495-4501.
10. Zhang, L.; Chen, M.; Wu, X.; Wang, H.; E, W.; Car, R. Deep neural network for the dielectric response of insulators. *arXiv e-prints*.
11. Unke, O. T.; Meuwly, M. PhysNet: A Neural Network for Predicting Energies, Forces, Dipole Moments, and Partial Charges. *J. Chem. Theory Comput.* **2019,** *15* (6), 3678-3693.
12. Medders, G. R.; Paesani, F. Infrared and Raman Spectroscopy of Liquid Water through "First-Principles" Many-Body Molecular Dynamics. *J. Chem. Theory Comput.* **2015,** *11* (3), 1145-1154.
13. Sommers, G. M.; Calegari Andrade, M. F.; Zhang, L.; Wang, H.; Car, R. Raman spectrum and polarizability of liquid water from deep neural networks. *Phys. Chem. Chem. Phys.* **2020,** *22* (19), 10592-10602.
14. Aidas, K.; Angeli, C.; Bak, K. L.; Bakken, V.; Bast, R.; Boman, L.; Christiansen, O.; Cimiraglia, R.; Coriani, S.; Dahle, P.; Dalskov, E. K.; Ekström, U.; Enevoldsen, T.; Eriksen, J. J.; Ettenhuber, P.; Fernández, B.; Ferrighi, L.; Fliegl, H.; Frediani, L.; Hald, K.; Halkier, A.; Hättig, C.; Heiberg, H.; Helgaker, T.; Hennum, A. C.; Hettema, H.; Hjertenæs, E.; Høst, S.; Høyvik, I.-M.; Iozzi, M. F.; Jansík, B.; Jensen, H. J. A.; Jonsson, D.; Jørgensen, P.; Kauczor, J.; Kirpekar, S.; Kjærgaard, T.; Klopper, W.; Knecht, S.; Kobayashi, R.; Koch, H.; Kongsted, J.; Krapp, A.; Kristensen, K.; Ligabue, A.; Lutnæs, O. B.; Melo, J. I.; Mikkelsen, K. V.; Myhre, R. H.; Neiss, C.; Nielsen, C. B.; Norman, P.; Olsen, J.; Olsen, J. M. H.; Osted, A.; Packer, M. J.; Pawlowski, F.; Pedersen, T. B.; Provasi, P. F.; Reine, S.; Rinkevicius, Z.; Ruden, T. A.; Ruud, K.; Rybkin, V. V.; Sałek, P.; Samson, C. C. M.; de Merás, A. S.; Saue, T.; Sauer, S. P. A.; Schimmelpfennig, B.; Sneskov, K.; Steindal, A. H.; Sylvester-Hvid, K. O.; Taylor, P. R.; Teale, A. M.; Tellgren, E. I.; Tew, D. P.; Thorvaldsen, A. J.; Thøgersen, L.; Vahtras, O.; Watson, M. A.; Wilson, D. J. D.; Ziolkowski, M.; Ågren, H. The Dalton quantum chemistry program system. *WIREs Computational Molecular Science* **2014,** *4* (3), 269-284.

15. Frisch, M. J.; Trucks, G. W.; Schlegel, H. B.; Scuseria, G. E.; Robb, M. A.; Cheeseman, J. R.; Scalmani, G.; Barone, V.; Petersson, G. A.; Nakatsuji, H.; Li, X.; Caricato, M.; Marenich, A. V.; Bloino, J.; Janesko, B. G.; Gomperts, R.; Mennucci, B.; Hratchian, H. P.; Ortiz, J. V.; Izmaylov, A. F.; Sonnenberg, J. L.; Williams; Ding, F.; Lipparini, F.; Egidi, F.; Goings, J.; Peng, B.; Petrone, A.; Henderson, T.; Ranasinghe, D.; Zakrzewski, V. G.; Gao, J.; Rega, N.; Zheng, G.; Liang, W.; Hada, M.; Ehara, M.; Toyota, K.; Fukuda, R.; Hasegawa, J.; Ishida, M.; Nakajima, T.; Honda, Y.; Kitao, O.; Nakai, H.; Vreven, T.; Throssell, K.; Montgomery Jr., J. A.; Peralta, J. E.; Ogliaro, F.; Bearpark, M. J.; Heyd, J. J.; Brothers, E. N.; Kudin, K. N.; Staroverov, V. N.; Keith, T. A.; Kobayashi, R.; Normand, J.; Raghavachari, K.; Rendell, A. P.; Burant, J. C.; Iyengar, S. S.; Tomasi, J.; Cossi, M.; Millam, J. M.; Klene, M.; Adamo, C.; Cammi, R.; Ochterski, J. W.; Martin, R. L.; Morokuma, K.; Farkas, O.; Foresman, J. B.; Fox, D. J. *Gaussian 16 Rev. B.01*, Wallingford, CT, 2016.
16. Berman, H. M.; Westbrook, J.; Feng, Z.; Gilliland, G.; Bhat, T. N.; Weissig, H.; Shindyalov, I. N.; Bourne, P. E. The Protein Data Bank. *Nucleic Acids Res.* **2000,** *28* (1), 235-242.
17. Westermayr, J.; Gastegger, M.; Menger, M.; Mai, S.; Gonzalez, L.; Marquetand, P. Machine learning enables long time scale molecular photodynamics simulations. *Chem. Sci.* **2019,** *10* (35), 8100-8107.
18. Abramavicius, D.; Palmieri, B.; Mukamel, S. Extracting single and two-exciton couplings in photosynthetic complexes by coherent two-dimensional electronic spectra. *Chem. Phys.* **2009,** *357* (1), 79-84.
19. Abramavicius, D.; Jiang, J.; Bulheller, B. M.; Hirst, J. D.; Mukamel, S. Simulation Study of Chiral Two-Dimensional Ultraviolet Spectroscopy of the Protein Backbone. *J. Am. Chem. Soc.* **2010,** *132* (22), 7769-7775.
20. Kasha, M.; Rawls, H. R.; El-Bayoumi, M. A. The exciton model in molecular spectroscopy. *Pure Appl. Chem.* **1965,** *11* (3-4), 371-392.
21. Andersen, H. C. Molecular dynamics simulations at constant pressure and/or temperature. *J. Chem. Phys.* **1980,** *72*, 2384-2393.
22. McQuarrie, D. A. *Statistical Mechanics*. University Science Books: Sausalito, 2000.
23. Thomas, M.; Brehm, M.; Fligg, R.; Vohringer, P.; Kirchner, B. Computing vibrational spectra from ab initio molecular dynamics. *Phys. Chem. Chem. Phys.* **2013,** *15* (18), 6608-6622.
24. Cheng, B.; Engel, E. A.; Behler, J.; Dellago, C.; Ceriotti, M. Ab initio thermodynamics of liquid and solid water. *Proceedings of the National Academy of Sciences* **2019,** *116* (4), 1110-1115.
25. Zhang, Y.; Hu, C.; Jiang, B. Bridging the Efficiency Gap Between Machine Learned Ab Initio Potentials and Classical Force Fields. *arXiv e-prints*.
26. Stone, A. J. Transformation between cartesian and spherical tensors. *Mol. Phys.* **1975,** *29* (5), 1461-1471.

Table S1: Hyperparameters for depicting embedded density used in the T-EANN models for PDMs and polarizability tensors of $H_2O$.

| Hyperparameters | $L$ | $\Delta r_s$ (Å) | $\alpha$ (Å$^{-2}$) | $r_c$ (Å) |
|---|---|---|---|---|
| $H_2O$ | 0/1 | 0 | 1.4 | 3.5 |
| $H_2O$ | 0/1 | 0.38 | 1.4 | 3.5 |
| $H_2O$ | 0/1 | 0.76 | 1.4 | 3.5 |
| $H_2O$ | 0/1 | 1.14 | 1.4 | 3.5 |
| $H_2O$ | 0/1 | 1.52 | 1.4 | 3.5 |
| $H_2O$ | 0/1 | 1.90 | 1.4 | 3.5 |
| $H_2O$ | 0/1 | 2.28 | 1.4 | 3.5 |
| $H_2O$ | 0/1 | 2.66 | 1.4 | 3.5 |
| $H_2O$ | 0/1 | 3.04 | 1.4 | 3.5 |
| $H_2O$ | 0/1 | 3.42 | 1.4 | 3.5 |

Table S2: Hyperparameters for depicting embedded density used in the T-EANN models for PDMs and polarizability tensors of $(H_2O)_2$.

| Hyperparameters | $L$ | $\Delta r_s$ (Å) | $\alpha$ (Å$^{-2}$) | $r_c$ (Å) |
|---|---|---|---|---|
| $(H_2O)_2$ | 0/1 | 0 | 0.70 | 5.0 |
| $(H_2O)_2$ | 0/1 | 0.54 | 0.70 | 5.0 |
| $(H_2O)_2$ | 0/1 | 1.08 | 0.70 | 5.0 |
| $(H_2O)_2$ | 0/1 | 1.62 | 0.70 | 5.0 |
| $(H_2O)_2$ | 0/1 | 2.16 | 0.70 | 5.0 |
| $(H_2O)_2$ | 0/1 | 2.70 | 0.70 | 5.0 |
| $(H_2O)_2$ | 0/1 | 3.24 | 0.70 | 5.0 |
| $(H_2O)_2$ | 0/1 | 3.78 | 0.70 | 5.0 |
| $(H_2O)_2$ | 0/1 | 4.32 | 0.70 | 5.0 |
| $(H_2O)_2$ | 0/1 | 4.86 | 0.70 | 5.0 |

Table S3: Hyperparameters for depicting embedded density used in the T-EANN models for PDMs and polarizability tensors of $(H_5O_2)^+$.

| Hyperparameters | $L$ | $\Delta r_s$ (Å) | $\alpha$ (Å$^{-2}$) | $r_c$ (Å) |
|---|---|---|---|---|
| $(H_5O_2)^+$ | 0/1 | 0 | 0.48 | 6.0 |
| $(H_5O_2)^+$ | 0/1 | 0.64 | 0.48 | 6.0 |
| $(H_5O_2)^+$ | 0/1 | 1.28 | 0.48 | 6.0 |
| $(H_5O_2)^+$ | 0/1 | 1.92 | 0.48 | 6.0 |
| $(H_5O_2)^+$ | 0/1 | 2.56 | 0.48 | 6.0 |
| $(H_5O_2)^+$ | 0/1 | 3.20 | 0.48 | 6.0 |
| $(H_5O_2)^+$ | 0/1 | 3.84 | 0.48 | 6.0 |
| $(H_5O_2)^+$ | 0/1 | 4.48 | 0.48 | 6.0 |
| $(H_5O_2)^+$ | 0/1 | 5.12 | 0.48 | 6.0 |
| $(H_5O_2)^+$ | 0/1 | 5.76 | 0.48 | 6.0 |

Table S4: Hyperparameters for depicting embedded density used in the T-EANN models for PDMs and polarizability tensors of liquid water.

| Hyperparameters | $L$ | $\Delta r_s$ (Å) | $\alpha$ (Å$^{-2}$) | $r_c$ (Å) |
|---|---|---|---|---|
| Liquid water | 0/1 | 0 | 0.70 | 5.0 |
| Liquid water | 0/1 | 0.54 | 0.70 | 5.0 |
| Liquid water | 0/1 | 1.08 | 0.70 | 5.0 |
| Liquid water | 0/1 | 1.62 | 0.70 | 5.0 |
| Liquid water | 0/1 | 2.16 | 0.70 | 5.0 |
| Liquid water | 0/1 | 2.70 | 0.70 | 5.0 |
| Liquid water | 0/1 | 3.24 | 0.70 | 5.0 |
| Liquid water | 0/1 | 3.78 | 0.70 | 5.0 |
| Liquid water | 0/1 | 4.32 | 0.70 | 5.0 |
| Liquid water | 0/1 | 4.86 | 0.70 | 5.0 |

Table S5: Hyper-parameters for depicting embedded density used in the T-EANN model for TDM of N-methylacetamide (NMA) molecule for nπ*/ππ* excitation.

| Hyper-parameters | $L$ | $\Delta r_s$ (Å) | $\alpha$ (Å$^{-2}$) | $r_c$ (Å) |
|---|---|---|---|---|
| nπ*/ππ* | 0/1/2 | 0 | 0.70 | 6.0 |
| nπ*/ππ* | 0/1/2 | 0.53 | 0.70 | 6.0 |
| nπ*/ππ* | 0/1/2 | 1.06 | 0.70 | 6.0 |
| nπ*/ππ* | 0/1/2 | 1.59 | 0.70 | 6.0 |
| nπ*/ππ* | 0/1/2 | 2.12 | 0.70 | 6.0 |
| nπ*/ππ* | 0/1/2 | 2.65 | 0.70 | 6.0 |
| nπ*/ππ* | 0/1/2 | 3.18 | 0.70 | 6.0 |
| nπ*/ππ* | 0/1/2 | 3.71 | 0.70 | 6.0 |
| nπ*/ππ* | 0/1/2 | 4.24 | 0.70 | 6.0 |
| nπ*/ππ* | 0/1/2 | 4.77 | 0.70 | 6.0 |
| nπ*/ππ* | 0/1/2 | 5.30 | 0.70 | 6.0 |
| nπ*/ππ* | 0/1/2 | 5.83 | 0.70 | 6.0 |

Table S6: Absolute RMSEs for predicting polarizability tensor (in atomic unit) in several molecular systems with different implementations (see text for details).

| System | $\boldsymbol{\alpha}^{NN1}$ | $\boldsymbol{\alpha}^{NN1}+\boldsymbol{\alpha}^{NN2}$ | $\boldsymbol{\alpha}^{NN1}+\boldsymbol{\alpha}^{NN3}$ | $\boldsymbol{\alpha}^{NN1'}+\boldsymbol{\alpha}^{NN2}+\boldsymbol{\alpha}^{NN3}$ | $\boldsymbol{\alpha}^{NN1}+\boldsymbol{\alpha}^{NN2}+\boldsymbol{\alpha}^{NN3}$ |
|---|---|---|---|---|---|
| $H_2O$ | 9.6 | 9.6 | $3.6\times10^{-1}$ | $1.0\times10^{-3}$ | $9.6\times10^{-4}$ |
| $(H_2O)_2$ | 8.4 | 7.8 | $9.1\times10^{-1}$ | $5.4\times10^{-1}$ | $4.3\times10^{-1}$ |
| $(H_5O_2)^+$ | 0.95 | 0.83 | $7.9\times10^{-2}$ | $5.1\times10^{-2}$ | $2.4\times10^{-2}$ |

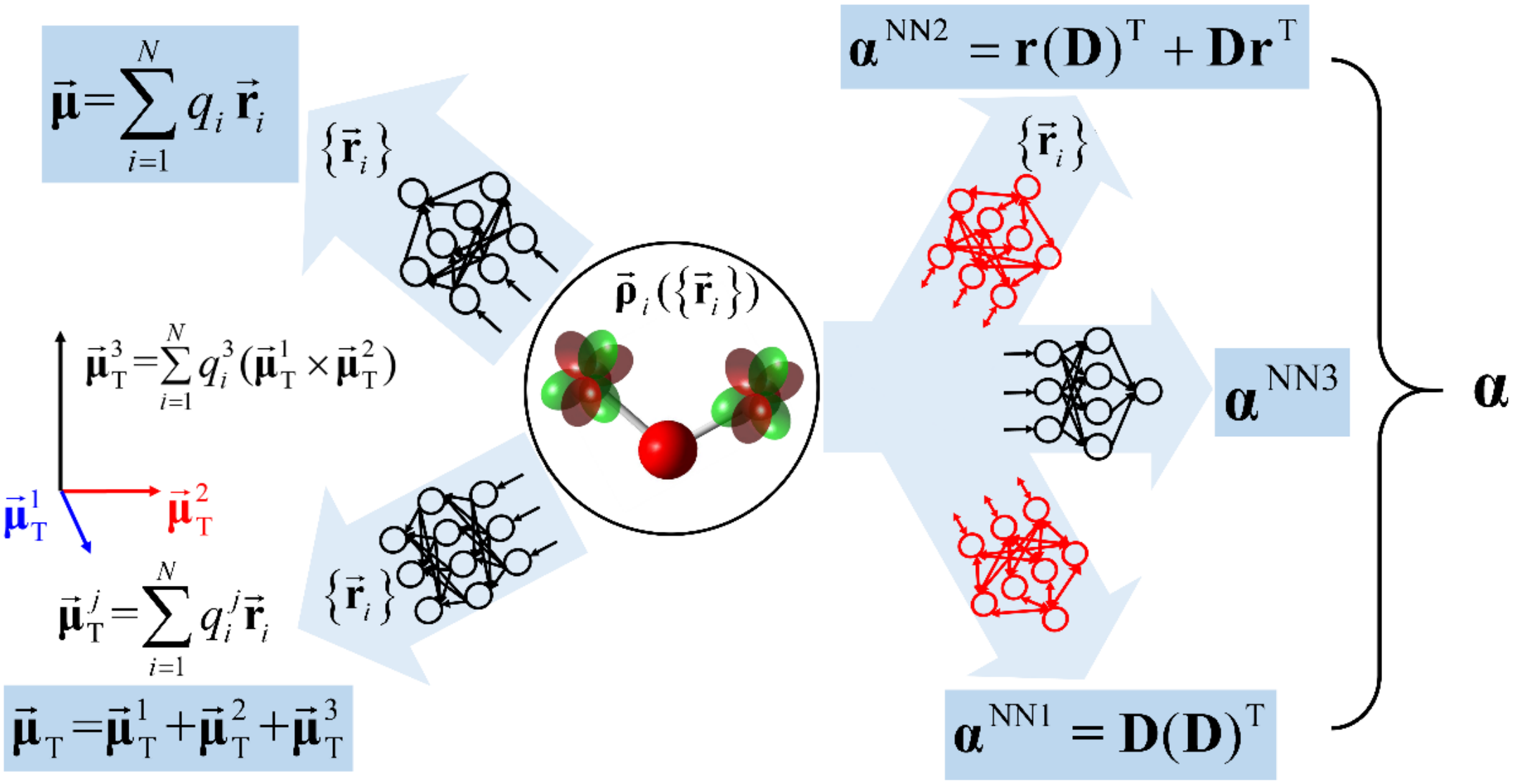

Fig. S1 Schematic diagram of tensorial embedded atom neural network models for PDM, TDM, and polarizability tensor. Regular feedforward neural networks are shown in black while the red ones indicate that partial derivatives are evaluated.

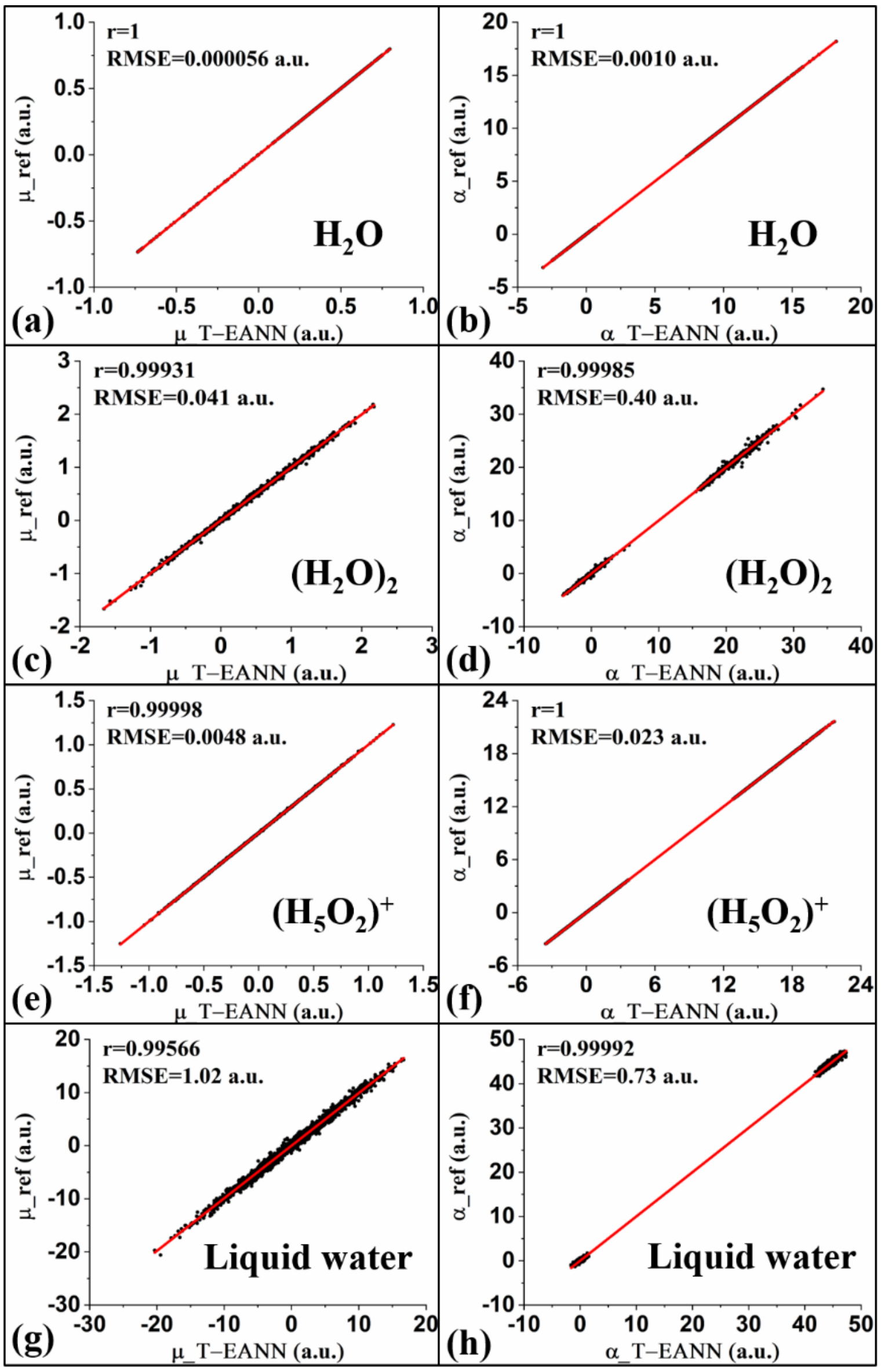

Fig. S2 Scatter plots of individual ab initio and T-EANN values (in a.u.) of dipole moment (left panel) and polarizability (right panel) of $H_2O$, $(H_2O)_2$, $(H_5O_2)^+$ and liquid water.

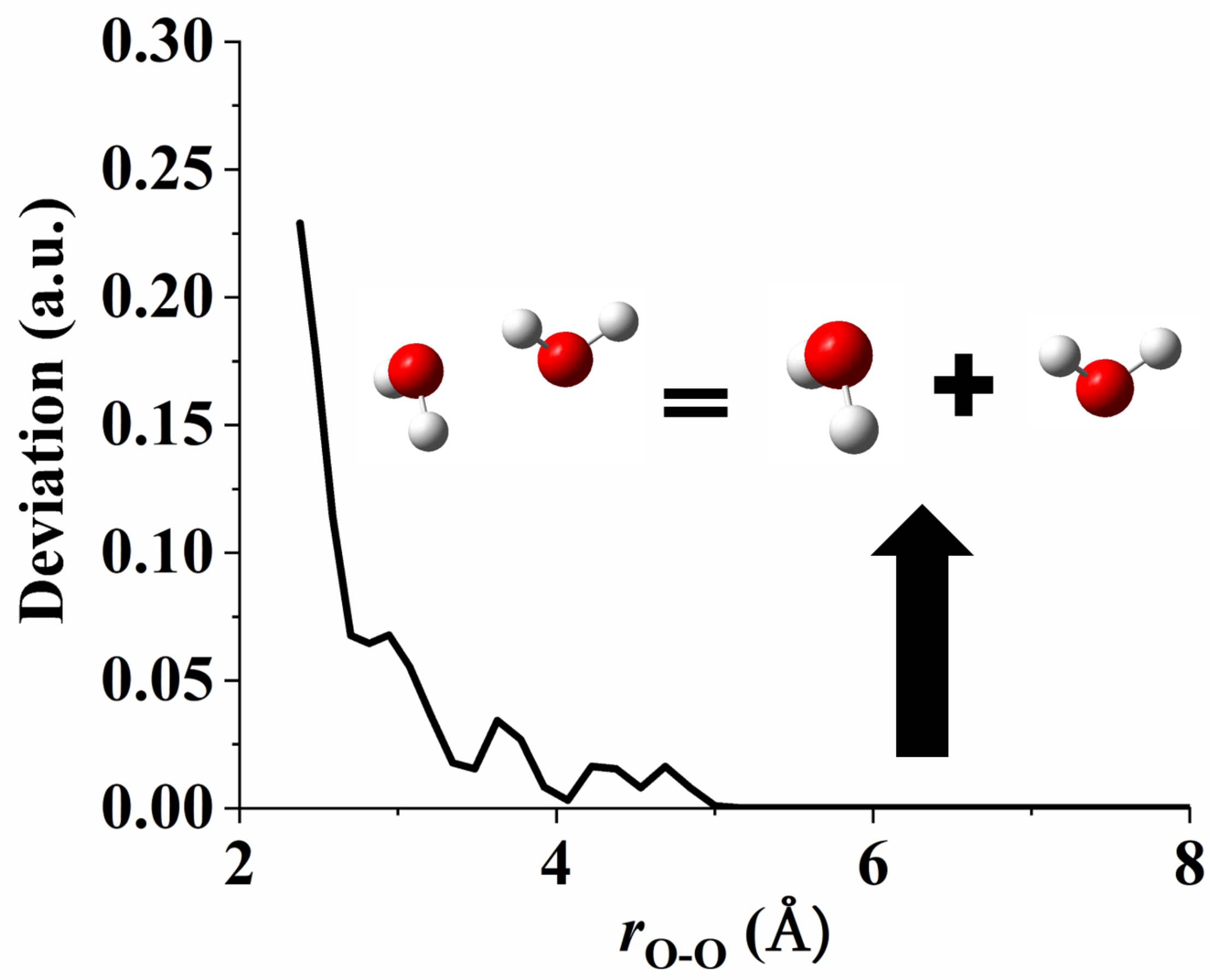

Fig. S3 Difference between permanent dipole moment of two monomer contributions and a water dimer, predicted by the T-EANN model, as a function of O-O distance. This Figure shows that the dipole moment of the water dimer separates rigorously in the limit of infinite separation of the two monomers. Similar behavior is found for polarizability.